\documentclass[aps,prl,showpacs,twocolumn,superscriptaddress]{revtex4}

\usepackage{graphicx,t1enc}

\begin{document}

\title{Analytic Considerations in the Study of Spatial Patterns Arising from Non-local Interaction Effects in Population
Dynamics}
\author{M. A. Fuentes}
\affiliation{Consortium of the Americas for Interdisciplinary
Science and Department of Physics and Astronomy, University of
New Mexico, Albuquerque, NM 87131, U.S.A.}
\author{M. N. Kuperman}
\affiliation{Consortium of the Americas for Interdisciplinary
Science and Department of Physics and Astronomy, University of
New Mexico, Albuquerque, NM 87131, U.S.A.} \affiliation{Centro
At{\'o}mico Bariloche and Instituto Balseiro, 8400 S. C. de
Bariloche, Argentina}
\author{V. M. Kenkre}
\affiliation{Consortium of the Americas for Interdisciplinary
Science and Department of Physics and Astronomy, University of
New Mexico, Albuquerque, NM 87131, U.S.A.}

\begin{abstract}

\ \newline

\ \newline

Simple analytic considerations are applied to recently discovered
patterns in a generalized Fisher equation for population
dynamics. The generalization consists of the inclusion of
non-local competition interactions among individuals. We first
show how stability arguments yield a condition for pattern
formation involving the ratio of the pattern wavelength and the
effective diffusion length of the individuals. We develop a
mode-mode coupling analysis which might be useful in shedding some light on the observed
formation of small-amplitude versus large-amplitude patterns.

\end{abstract}

\pacs{87.17.Aa, 87.17.Ee, 87.18.Hf}
\maketitle

We have shown recently that the Fisher equation used frequently
for investigations of biological or ecological systems, when
generalized to include spatially non-local competition
interactions, leads to interesting patterns in the steady state
density \cite{fkk,vmkpasi}. In this Note we attempt to shed some
analytic light on the formation of these patterns. The original
Fisher equation \cite{fish,murray} is
\begin{equation}
\frac{\partial u\left(\vec{x},t\right) }{\partial t} =D\nabla
^{2}u\left(\vec{x},t\right) +au\left( \vec{x},t\right)
-bu^{2}\left(\vec{x},t\right), \label{originaleq}
\end{equation}
where $u\left(\vec{x},t\right)$ is the population density of
individuals under investigation (bacteria, rodents, etc.) at
position $\vec{x}$ and time $t$, and $D$, $a$, $b$ are,
respectively, the diffusion coefficient, population growth rate,
and competition parameter. The generalized equation
\cite{fkk,vmkpasi} features competition interactions linking
$u(\vec{x},t)$ at point $\vec{x}$ with $u(\vec{y},t)$ at point
$\vec{y}$ through an influence function $f_{\sigma
}(\vec{x},\vec{y})$ of range $\sigma$,

\begin{eqnarray}
\frac{\partial u\left( \vec{x},t\right) }{\partial t} &=&D\nabla
^{2}u\left(\vec{x},t\right) +a\,u(\vec{x},t)  \label{influeq} \\
&&-b\,u(\vec{x},t)\int_{\Omega }u(\vec{y},t)f_{\sigma
}(\vec{x},\vec{y})dy, \nonumber
\end{eqnarray}
$\Omega$ being the domain for the non-local interaction.

In \cite{fkk}, we found that the introduction of the finite-range
competition interactions gives rise to the emergence of patterns
in the steady state density $u(\vec{x})$ with the following
features:
\begin{itemize}
  \item No patterns appear \cite{vmkpasi} in the two extremes of zero range (in which the generalization reverts to the Fisher equation) and full range (in which the population density is linked equally to all points in the domain).
  \item The pattern structure depends crucially on features of the influence function, specifically, its cut-off length and its width.
  \item Even when patterns appear, their steady-state amplitude can change abruptly from substantial to negligible as the parameters of the system are varied.
  \item The critical quantity determining the separation of large-amplitude patterns from small-amplitude ones appears to be  the ratio of the cut-off length of the influence function to its width.
\end{itemize}

These findings raise two questions. Why do the patterns form at
all? And, what causes the separation of the large-amplitude
patterns from  the small-amplitude ones? Both questions are
interesting. The first is amenable to understanding via standard stability
analysis considerations.  The second is more difficult but might be
approachable through a mode-mode coupling analysis, as we show
below.

In order to address the first question, consider a 1-dimensional
version of Eq. (\ref{influeq}) for simplicity, and substitute in it
\begin{equation}
u(x,t) =u_{0}+\epsilon \cos (kx)\exp (\varphi t). \label{subs}
\end{equation}
Here $u_{0}$ is the homogeneous steady-state solution $a/b$.
Considering periodic boundary conditions, and retaining only
first order terms in $\epsilon$, we obtain the following
dispersion relation between the wavenumber $k$ of any mode of the
pattern and the rate $\varphi$ at which it tends to grow:
\begin{equation}
\varphi =-Dk^{2}-a\mathcal{F}(k).  \label{alge}
\end{equation}
In this expression, the influence function (assumed to be even)
is represented by its  cosine (Fourier) transform
$\mathcal{F}(k)$ defined as
\begin{equation}
\mathcal{F}(k)=\int_{\Omega }\cos (kz)f_{\sigma }(z)dz.
\end{equation}
Stable steady-state patterns require that
\begin{equation}
\lambda >2\pi \sqrt{\frac{D}{-a\mathcal{F}(\lambda )}},
\label{condinest}
\end{equation}
where $\lambda=2\pi /k$ is the wavelength associated with the
$k-$mode of the Fourier expansion of the pattern.

Condition (\ref{condinest}) allows us to check for the existence
or absence of inhomogeneity, i.e., patterns, in the steady state. We see
from (\ref{condinest}) that the Fourier transform of the
influence function at the wavelength under consideration should
be \emph{negative} for the patterns to appear and that its
magnitude should be large enough. One way of understanding this
condition is to recast it as requiring that  $2\pi$ times the
`effective diffusion length'  should be smaller than the
wavelength for the patterns to occur. By the effective diffusion
constant is meant $D$ divided by $-{\mathcal F}(\lambda)$, which is a factor decided by the influence function, and by the diffusion length is
meant the distance traversed diffusively in a time interval of
the order of the inverse of the growth rate. If the influence
function is smooth such as in the case of a Gaussian in an
infinite domain, the Fourier transform is positive and no
patterns appear. A cut-off in the influence function produces
oscillations in the Fourier transform which can go negative for
certain wavelengths. The reported finding \cite{fkk} that the
cut-off nature is essential to pattern formation can be
understood naturally in this way.

Let us consider, in turn, three cases of the influence function
which we have used in our earlier investigations \cite{fkk}:
square, cut-off Gaussian, and intermediate.

First we take
\begin{equation}
f(x-y)=\frac{1}{2w} \{\theta [w-(x-y)]\theta [w+(x-y)]\},
\label{influsq}
\end{equation}
where $\theta$ is the Heaviside function. The influence function
is thus a square of cut-off range measured by $w$ from its center.
We will consider the case here when the range $w$ is smaller than, or equal to, the domain length $L$. Equation (\ref{alge}) then involves an integral from $0$ to $w$, and gives
\begin{equation}
\varphi =-a\frac{\sin (kw)}{kw}-Dk^{2}.  \label{omega}
\end{equation}
In terms of dimensionless parameters
\begin{eqnarray}
\varphi ^{\prime } &=&\varphi /a, \nonumber \\
k^{\prime } &=&k\sqrt{D/a}, \nonumber \\
\eta &=&w \sqrt{a/D}, \nonumber
\end{eqnarray}
we have
\begin{equation}
\varphi ^{\prime }=-\frac{\sin (k^{\prime }\eta )}{k^{\prime
}\eta }-k^{\prime 2}, \label{dimless}
\end{equation}
which we plot in Fig. 1 for three different values ($50$, $10$
and $2$) of the ratio $\eta$ of the width to the diffusion length
(not \emph{effective} diffusion length). For the third case there are no
patterns: diffusion is strong enough to wash them out. For the
intermediate case, patterns can occur with wavelengths
corresponding to values around $k' \approx 0.4$ while for the
$\eta=50$ case, they occur around $k' \approx 0.1$.

\begin{figure}
\includegraphics[width=7cm,angle=-90]{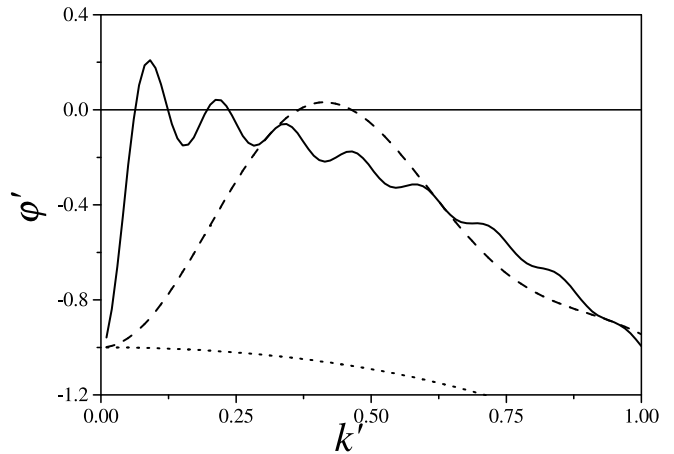}\\
\caption{The dispersion relation (\ref{dimless}) between the
dimensionless growth exponent $\varphi ^{'}$ and wavenumber $k
^{'}$ plotted  for different values of the ratio $\eta$ of the
influence function range to the diffusion length (see text).
Values of $\eta$ are 50 (solid line), 10 (dashed line), and  2
(dotted line). Patterns appear for those values of $k'$ for which
$\varphi$ is positive.} \label{f1}
\end{figure}

The earlier finding \cite{fkk,vmkpasi} that no patterns appear
for extremes of the range of the influence function is clear from
Eq. (\ref{dimless}).  As the influence width vanishes, i.e., as
$\eta$ goes to zero, both terms in $\varphi$ are negative and
there can be no steady state patterns:  we recover the solution
for the local limit, corresponding to Eq. (\ref{originaleq}),
when $w\rightarrow 0$. Since the boundary conditions are periodic
in a domain of length $L$, there are only the allowed values $k=n
\pi/L$ of the wavenumber. Therefore, in the opposite limit of
full range, i.e., $w\rightarrow L$, the sine term vanishes,
$\varphi ^{\prime }=-k^{\prime 2}$, and again there are no
patterns.

Precisely the same qualitative behavior occurs for other
non-square influence functions such as the Gaussian with a cut-off,
i.e., for
\begin{equation}
f(x-y)=\frac{1}{\sigma \sqrt{\pi}\mbox{erf} ( w/\sigma)} exp\left[ \left( \frac{x-y}{\sigma} \right)^2
\right].
\end{equation}
We again consider the case when the cut-off length does not exceed the domain length. This leads to the Fourier transform of the influence function involving an integral from $0$ to $w$. In these as well as other cases considered, it should be appreciated that the domain length $L$, if taken to be smaller than the cut-off length, becomes itself the cut-off length: factors such as $kw$ appearing in the Fourier transform become then $kL$ instead.

For this cut-off Gaussian case, the square case dispersion
relation (\ref{omega}) is replaced by
\begin{eqnarray}
\varphi &=&-\frac{a\exp [-(k\sigma /2)^{2}]}{2\mbox{erf}(w/\sigma)} \nonumber \\
&&\left[ \mbox{erf}\left(\frac{w}{\sigma}-\frac{ik\sigma}{2}\right)+\mbox{erf}\left(%
\frac{w}{\sigma }+\frac{ik\sigma}{2}\right)\right] -Dk^{2}.
\nonumber \label{gaussdisp}
\end{eqnarray}
The dimensionless version (\ref{dimless}) is replaced by
\begin{equation}
\varphi ^{\prime }=-\frac{\exp (-k^{\prime 2}\beta
^{2})}{2\mbox{erf}(\alpha )}\left\{ \mbox{erf}(\alpha -ik^{\prime
}\beta )+\mbox{erf}(\alpha +ik^{\prime }\beta )\right\}
-k^{\prime 2}. \nonumber
\end{equation}
Here $\alpha$ and $2\beta$ are the ratios of the cut-off length to the range and of the range to the diffusion length respectively:
\begin{eqnarray}
\alpha &=&w/\sigma, \nonumber \\
2\beta &=&\sigma \sqrt{a/D}. \nonumber
\end{eqnarray}
What is analogous to $\eta$ in the square case is their product $2\alpha \beta=w\sqrt{a/D}$. Plots which are essentially the same as those in Fig. 1 can be drawn for this Gaussian case.

It is interesting to note that, while there is a single quantity
$\eta$ in the square case, there are two quantities,  $\alpha$
and $\beta$, in the cut-off Gaussian case. This arises from the
fact that, although there are generally two lengths associated
with any influence function, the cut-off length and the width,
the latter is infinite for the square case. The width has been
defined in Ref. \cite{fkk} as being inversely proportional to
the  second derivative of the influence function evaluated at its
central point, and has been denoted by the symbol $\Sigma$. The
cut-off length measures the distance beyond which the influence
function is exactly zero and has been denoted \cite{fkk} by
$\xi_c$. For the cut-off Gaussian, this $\xi_c=x_c=w$. The symbol
$x_c$ has been used in Ref. \cite{fkk} and $w$ in the present
Note. The width $\Sigma$ obeys $\Sigma=\sigma$ for the Gaussian
case and $\Sigma=\infty$ for the square case.

Both the cut-off Gaussian and the square can be obtained as
particular cases of a general function \cite{fkk,tsa}  which we call the \emph{intermediate} influence function:
\begin{eqnarray}
f(x-y)&=& \frac{\Gamma (1/r+3/2)}{\sqrt{\pi }w\Gamma (1/r+1)}
\left [ 1-\frac{r\left(x-y\right)^{2}}{(2+3r) \sigma^{2}} \right ]^{1/r} \nonumber \\
& & \{\theta [w-(x-y)]\theta [w+(x-y)]\}.
\end{eqnarray}

For $r \rightarrow 0$ the intermediate influence function goes to the cut-off
Gaussian, while for $r \rightarrow \infty$ it yields he square
function. The cut-off length of the influence
function is given  by
\begin{equation}
w=\sqrt{\frac{(2+3r)}{r}}\sigma. \nonumber
\end{equation}
We will follow the notation
\begin{eqnarray*}
\nu&=&1/r +1/2,\nonumber
\end{eqnarray*}
and evaluate the Fourier transform of the influence function by calculating the integral  \cite{Ryzhuk}
\begin{eqnarray}
&&\frac{2 w^{-2 \nu}\Gamma (\nu+1)}{\sqrt{\pi }\Gamma (\nu+ \frac{1}{2})} \int_{0}^{w}\cos (ks) \left[ w^{2}-s^{2}\right] ^{\nu -\frac{1}{2}}ds\nonumber \\
&=&\left( \frac{2}{k w}\right)^{\nu} \Gamma(\nu+1)J_{\nu}(kw), \nonumber \\
\end{eqnarray}
for
\[
k>0,w>0,\mbox{Re}[\nu ]>\frac{1}{2},
\]
$\Gamma$ and $J$ being the gamma and the Bessel functions respectively. The dimensionless dispersion relation analogous to (\ref{dimless}) is, for this general case,
\begin{equation}
\varphi ^{\prime}
=-\left( \frac{2}{k^{\prime }\eta}\right) ^{\nu }\Gamma (\nu
+1)J_{\nu }(k^{\prime } \eta )-k^{\prime 2}.
\label{drtsallis} \\
\end{equation}
Here, as in the square case, $\eta=w\sqrt{a/D}$ is the ratio of the influence function width to the diffusion length.

\begin{figure}
\includegraphics[width=7cm,angle=-90]{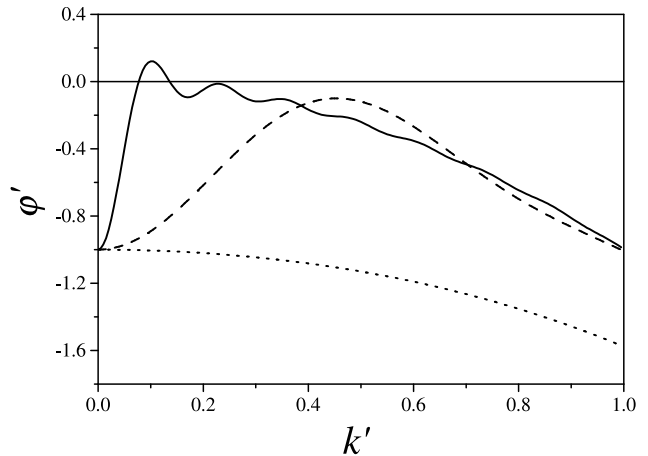}\\
\caption{The dispersion relation (\ref{drtsallis}) between the
dimensionless growth exponent $\varphi ^{'}$ and wavenumber $k
^{'}$ plotted  for the \emph{intermediate} influence function. Values of $\eta$ are as in Fig. 1: 50 (solid line), 10 (dashed line), and  2
(dotted line). Patterns appear for those values of $k'$ for which
$\varphi ^{'}$ is positive.}
\label{f2}
\end{figure}

It is straightforward to obtain the two limits, square and Gaussian, from this  dispersion result (\ref{drtsallis}) for the intermediate influence function. In Fig. 2 we plot the intermediate case for $\nu=1$ and see the same general behavior as in the Gaussian and the square counterparts (see, e.g., Fig. 1). Steady state patterns appear only around $k'=0.1$.

Having understood the appearance of the patterns, we now come to the second issue  mentioned in the introduction: the transition from small-amplitude to large-amplitude patterns
\cite{fkk}. This is much more difficult to address analytically. We present here a procedure that we believe has the potential to  shed
some light on this issue. We substitute the Fourier mode expansion of  $u(x,t)$,
\begin{equation}
u\left( \vec{x},t\right) =\sum A_{n}(t)\cos (k_{n}x), \nonumber
\end{equation}
in (\ref{influeq}), explicitly noting that $k_{n}=\pi n/L$, and
using the orthogonality properties of trigonometric functions,
obtain separate equations for the $n=0$ mode,
\begin{eqnarray}
\frac{dA_{0}}{dt} &=&aA_{0}-bA_{0}^{2}-b\,\sum_{n=1}^{\infty
}\frac{A_{n}^{2}}{2}\mathcal{F}(k_{n}),\label{amplitude0}
\end{eqnarray}
and for other modes $n\neq 0$:
\begin{eqnarray}
\frac{dA_{n}}{dt} &=&-Dk_{n}^{2}A_{n}+aA_{n}  \label{amplituden} \\
&&-bA_{0}A_{n}\left[ 1+\mathcal{F}(k_{n})\right]   \nonumber \\
&&-b\,\sum_{j=1}^{n-1}\frac{A_{j}A_{n-j}}{2}\mathcal{F}(k_{j})  \nonumber \\
&&-b\,\sum_{j=n+1}^{\infty }\frac{A_{j}A_{j-n}}{2}  \nonumber \\
&&\left[ \mathcal{F}(k_{j})+\mathcal{F}(k_{j-n})\right].
\nonumber
\end{eqnarray}
Equations (\ref{amplitude0}) and (\ref{amplituden})
are the complete set of equations for the evolution of the
amplitudes of all modes in the non-local problem given by Eq.
(\ref{influeq}). The appearance of  stable patterns only for
those values of $k_{n}$ for which $\varphi$ is positive as seen
in our Figs. 1 and 2, suggests
that we envisage an interaction between \emph{only two modes}, the zero
mode and the one whose growth we examine, say $n=m$. In a situation as in the plots shown in which $\varphi>0$ only for a small $k-$ range, the discrete nature of the allowed $k$ values could lead to only a single non-zero mode lying in the stable range. Then we would have only two coupled nonlinear equations for the mode amplitudes,
\begin{eqnarray}
\frac{dA_{0}}{dt} &=&aA_{0}-bA_{0}^{2} -b\,\frac{A_{m}^{2}}{2}\mathcal{F}(k_{m}),  \nonumber \\
\frac{dA_{m}}{dt} &=&-Dk_{m}^{2}A_{m}+aA_{m} -bA_{0}A_{m}\left[
1+\mathcal{F}(k_{m})\right] \nonumber
\end{eqnarray}
which lead, in the steady state, to
\begin{eqnarray}
A_{0} &=&\frac{a-Dk_{m}^{2}}{b\left[ 1+\mathcal{F}(k_{m})\right]
}   \\
A_{m}^{2}&=&-2\left[\frac{bA_{0}^{2}-aA_{0}}{b\mathcal{F}(k_{m})}\right].\label{an}\end{eqnarray}

Substitution of the zero mode amplitude in $A_m$ gives an explicit expression for the latter:
\begin{equation}
A_{m}^{2}=-\frac{2}{\mathcal{F}(k_m) b} \left[\frac{(a-D k_m^2)^2}{b
(1+\mathcal{F}(k_m))^2}-
\frac{a (a-D k_m ^2 )}{b (1+\mathcal{F}(k_m)}\right].
\label{twomode}
\end{equation}

Preliminary investigations lead us to believe that the appearance of small and large amplitude patterns for different parameter regimes might emerge from considerations of (\ref{twomode}). The two-mode approach is, however, plagued by the fact that variation of the parameters of the system can make the approach invalid in one regime even if it is valid in another. Surely steady state pattern amplitudes can always be obtained by a simultaneous solution of the algebraic equations obtained by putting the left side of (\ref{amplitude0}) and (\ref{amplituden}) to zero. This is a straightforward numerical task if we can make the reasonable assumption that the number of modes to be considered has a finite cut-off. Such a cut-off is an obvious consequence of the fact the Fourier transform of a typical influence function  disappears for high values of $n$. These and related studies will be reported in the future.

\section{Acknowledgements}

This work is supported in part by the Los Alamos National
Laboratory via a grant made to the University of New Mexico
(Consortium of the Americas for Interdisciplinary Science), by the
NSF's Division of Materials Research via grant No DMR0097204, by
the  NSF's International Division via grant No INT-0336343, and
by DARPA-N00014-03-1-0900.

\end{document}